\documentclass[a4paper,11pt]{article}
\pdfoutput=1 

\usepackage{jheppub} 

\usepackage{graphicx}
\usepackage{verbatim}
\usepackage{placeins} 
\usepackage{color}
\usepackage[normalem]{ulem} 
\usepackage{lmodern}
\usepackage{todonotes}		
\usepackage[utf8]{inputenc}
\usepackage{amsmath}
\usepackage{subcaption}

\newcommand{\njet}{$N^{Jet}$}
\newcommand{\htt}{$H_T$}

\newcommand{\hjm}{High~$N^{Jet}$}

\newcommand{\pt}{$p_T~$}

\newcommand{\tbt}{$2 \rightarrow 2$}

\newcommand{\njetf}{$N^A_{Jets}$}
\newcommand{\njets}{$N^B_{Jets}$}

\newcommand{\avnj}{$ \langle N^{Jet} \rangle $}

\newcommand{\xy}{$(x,y)$}

\author[1]{Daniel Turgeman,}
\author[2]{Michael Pitt,}
\author[1]{Itamar Roth,}
\author[1]{and Ehud Duchovni.}

\affiliation[1]{Weizmann Institute of Science, Rehovot 7610001, Israel}
\affiliation[2]{CERN, 1211 Geneva 23, Switzerland}

\title{\boldmath On the Modelling of  Energetic Multi-jet QCD Events}

\abstract
{Physics beyond the Standard Model (BSM) may be unveiled by studying events with a high number of outgoing jets, produced at the LHC with energies above the TeV scale (energetic multi-jet events). Such events are dominated by QCD processes, where the calculations rely on some sort of approximation. Therefore, it is important to develop a robust approach for modeling such events that could probe the existence of BSM signals.
In this note, jet spatial distributions in energetic multi-jet processes were compared using several state-of-the-art MC event generators. Slight differences were found, indicating modelling limitations. Therefore, a data-driven technique for the estimation of processes with a final state that contains a large number of jets is proposed. This procedure can predict jet multiplicities up to a precision of ~25\% in energetic multi-jet events.}

\begin{document}
	%
	\maketitle
	\flushbottom
\section{Introduction}\label{sect:intro}
Several \textit{beyond the standard model} (BSM) models, e.g   Micro-Black-Holes (mBH) \cite{Giddings:2001bu,Giddings:2001ih,Dimopoulos:2001hw,Chamblin:2002ad}, R-Parity Violation (RPV) Supersymmetry \cite{Dreiner:1997uz,Barbier:2004ez} or Sphaleron induced processes \cite{Ellis:2016ast,Papaefstathiou:2019djz} predict the possible production at the LHC of events with a large number of outcoming high energy partons . These events will give rise to final states consisting of a high multiplicity of jets, produced above the TeV scale, namely, energetic multi-jet events. \\
The identification of this type of signal, through the observation of an excess of energetic multi-jet events, is far from being straightforward due to the presence of large Standard Model (SM) background originating from Quantum Chromodynamics (QCD) processes. 
The presently available event generators for such processes perform Leading Order (LO), Next to leading order (NLO) or even partial NNLO calculations followed by radiation of additional partons through the Parton Shower (PS) algorithm. 
The accuracy of these calculations is limited as some unavoidable approximations must be imposed. 
Significant effort has been invested in the study of energetic multi-jet events at the LHC with the ATLAS and CMS collaborations \cite{Sirunyan:2018xwt,Mjet_13,Aaboud:2018lpl}. These studies cope with a major difficulty, namely, with the need to estimate the kinematics of multi-jet events without the usage of simulation. Yet some indirect dependence on simulation always remains.
As a first stage in the development of a novel data-driven technique for estimation of QCD processes, the difference between predictions of several event generators, when energetic  multi-jet events are simulated, is presently studied. The description of the technique, as well as its performance using the aforementioned simulations, are the subjects of the latter sections.

\section{Simulation of Multijet Processes} \label{generators}
Modeling multi-jet processes in QCD is a challenging task mainly due to the large number and high complexity of the relevant Feynman diagrams.  Generally speaking, there are three main approaches to handle this complexity:

The simplest approach is to couple the Leading Order (LO) calculations that give rise to two outcoming partons with a Parton Shower (PS) algorithm that can produce additional jets. While such an approach provides precise physical modelling of di-jet with soft-collinear parton emissions, it is less accurate pertaining to topologies with more than two well separated partons. In spite of its simplicity the LO+PS technique provides a surprisingly good description of the Tevatron and LHC data  \cite{Sirunyan:2018adt}.

A better simulation can be achieved by carrying	 out the calculations of additional real parton emissions, while neglecting the contribution of virtual corrections. Such an approach can give rise to events with up to four or five jets. In order to simulate higher jet multiplicities a PS algorithm is applied. While such an approach improves the description of final states with more than two well separated partons, the interfacing with a PS algorithm requires a proper matching scheme.

The most rigorous  approach is a formal order-by-order perturbative calculation, where each extra order includes diagrams with one more outcoming particle  and one more loop in the intermediate state. Next to Leading Order (NLO) calculations can compute the properties of up to 3 jets with one loop corrections. Calculations of higher orders are very resource consuming and are thus limited.

For the purpose of comparison between the various event-generators strategies, each approach was used to simulate QCD events at $\sqrt{s}=$ 13 TeV as outlined below. Jets were reconstructed using the \textit{anti-kt} algorithm \cite{Cacciari:2008gp} implemented in the FastJet 3.2.1 package \cite{Cacciari:2011ma} with a radius parameter value of R = 0.4. All jets were required to satisfy \pt $>$ 50 GeV and $|\eta|<2.8$. 

\begin{itemize}
\item For the LO+PS approach events were generated and showered with PYTHIA8.235 \cite{Sjostrand:2014zea}. To efficiently cover the large phase-space (from the GeV to TeV scales)  the sample was generated in slices of the reconstructed leading jet's $p_T$ with a constant number of events simulated in each slice.
The leading reconstructed jet kinematics is affected by initial and final state radiation (ISR/FSR).  Therefore, the generator level parameter of $\hat{p}_T^{min}$, i.e. - the cut for the minimum transverse momentum of the outgoing leading parton at generator level, has to be set to be lower than the leading jet  $p_T$ used in defining the slices. Optimization studies found that for a sample with the leading jet $P_T$ between $p_T^{MIN}$ and $p_T^{MAX}$ GeV a cut of $\hat{p}_T^{min} = \left(p_T^{MIN}/395\right)^3+\left(p_T^{MIN}/164\right)^2+\left(p_T^{MIN}/1.85\right)$ was most efficient in minimizing computation time.

\item For the multi-leg approach events were simulated with the MadGraph5\_aMC@NLO v2.6.3.2 \cite{Alwall:2014hca}  event generator using matrix elements calculations for up to four partons at leading order. Events were generated in slices of  the total sum of partonic \pt ($\hat{H}_T$) covering the entire energy range. The use of $\hat{H}_T$ for defining the slices greatly minimized computation time and was possible due to the multi-leg simulation at generator level.
The generated events were fed into PYTHIA8 where the PS algorithm has been applied to all partons, using the CKKW-L merging scheme \cite{Lonnblad:2001iq,Lonnblad:2011xx}, with a merging scale of 80 GeV. 

\item In the last case of full NLO calculations the POWHEG-BOX v2 framework \cite{Nason:2004rx,Frixione:2007vw,Alioli:2010xd} has been used to simulate di-jet and three-jet processes \cite{Alioli:2010xa,Kardos:2014dua}. For full coverage of the entire energy range, the sample was creating using 350 GeV slices of Powheg-$k_T^{born}$ (the $\hat{p}_T$ of the underlying Born diagram).  Events were showered with PYTHIA8 using the default Powheg NLO merging scheme. 

\end{itemize}

In all cases, the factorization and renormalization scales are set to $H_T/2$, and the CT14 \cite{Hou:2016sho} PDF-set was used. Parton shower and underlying event was used with the Monash 2013 tune \cite{Skands:2014pea}.

\section{Results} \label{results}
%
\subsection{Event Kinematics} 
\begin{figure}
	\centering
	\includegraphics*[width=12cm]{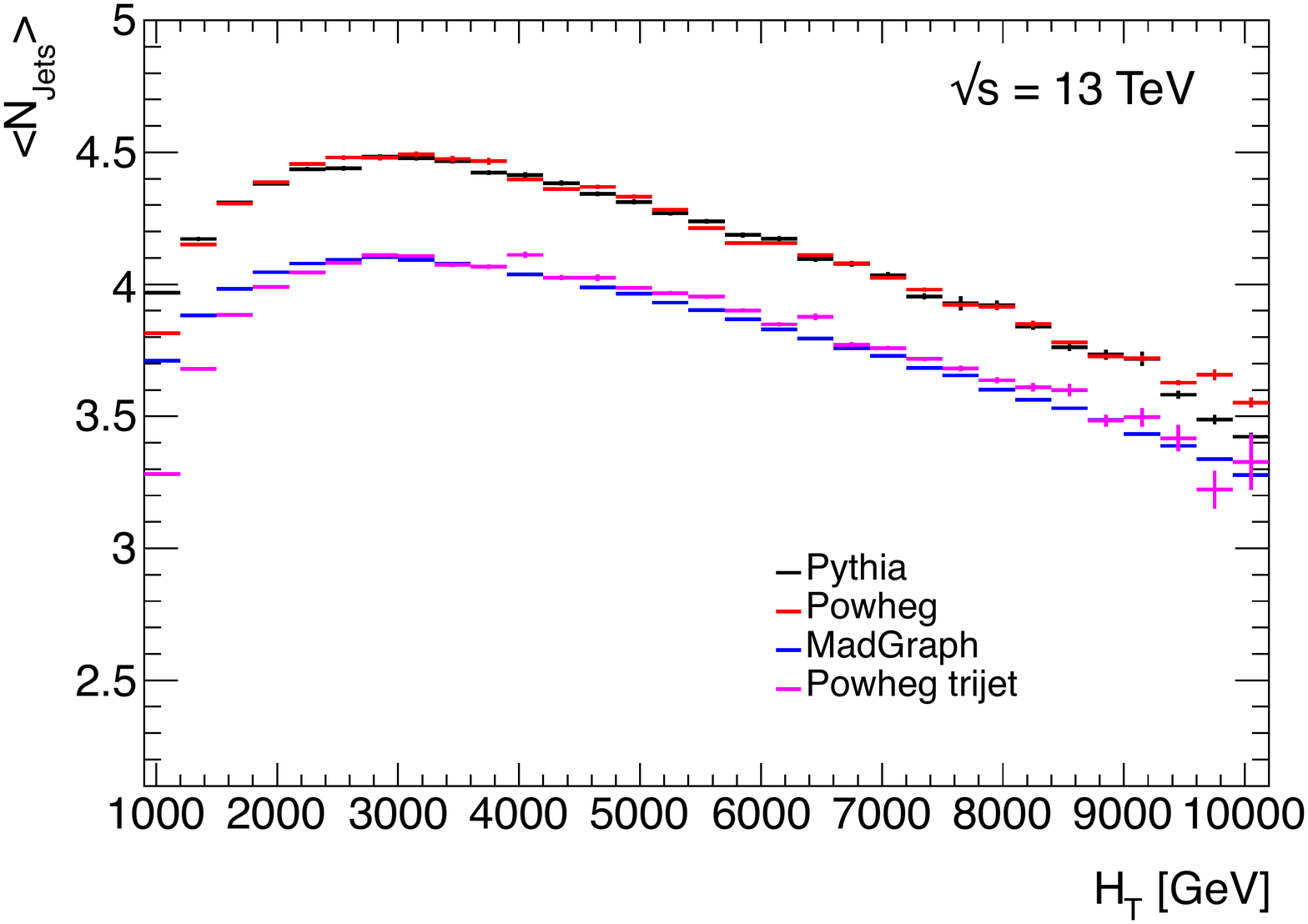}
	\caption{The average number of jets per event (\avnj) as a function of \htt. The result is stable under a change of the minimal transverse momentum and maximal pseudo-rapidity for jet acceptance. The  lower average jet multiplicty exhibited by Madgraph and Powheg trijet may be attributed to the QCD scale uncertainties (not shown) and requires further investigation.}
	\label{njvsht}
\end{figure}

\begin{figure}
	\centering
	\includegraphics*[width=12cm]{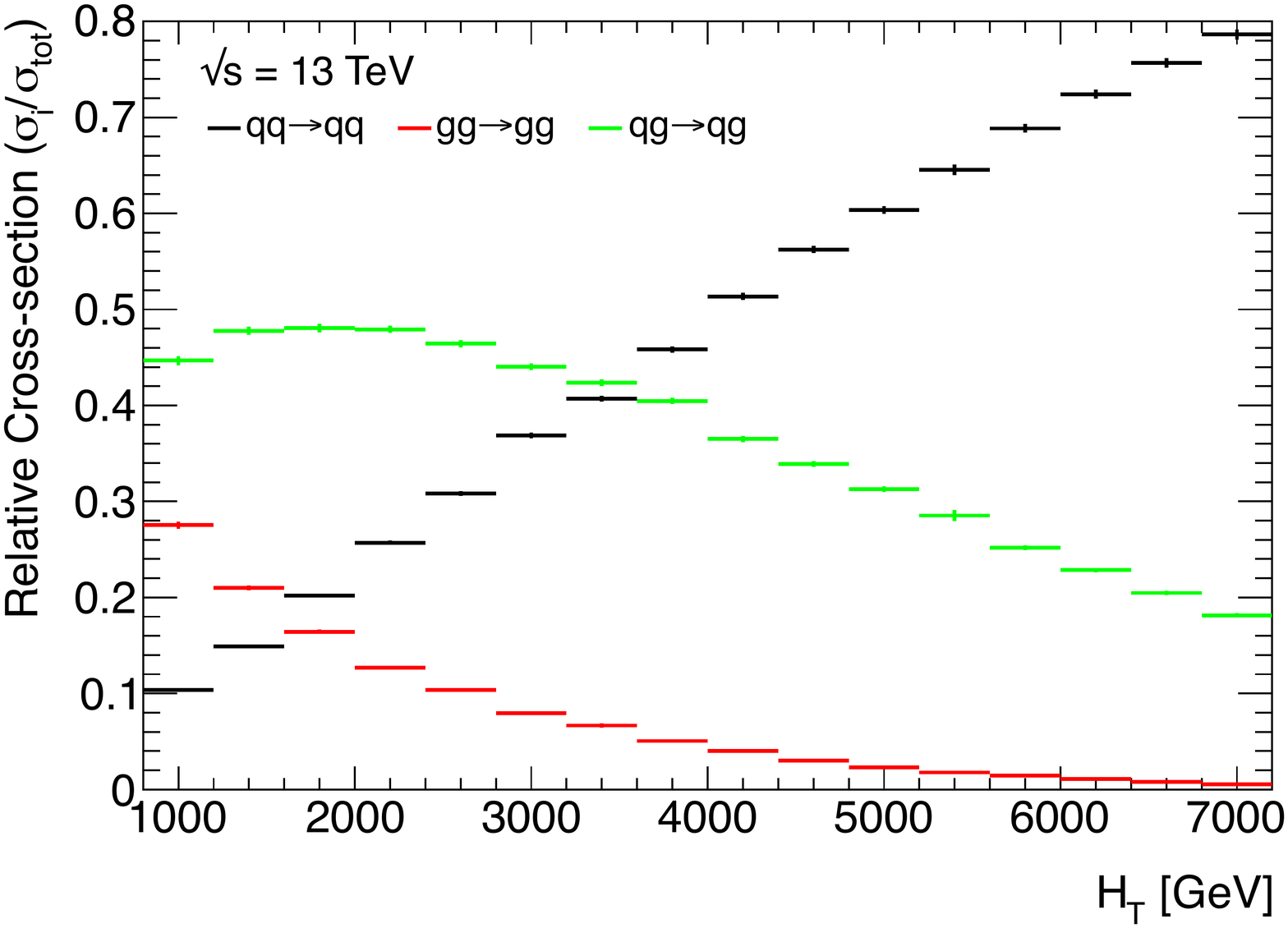}
	\caption{The relative cross-sections at LO of the 3 leading subprocesses. Events were generated using Pythia.} 
	\label{subprocess}
\end{figure}

\begin{figure}
	\centering
	\includegraphics*[width=12cm]{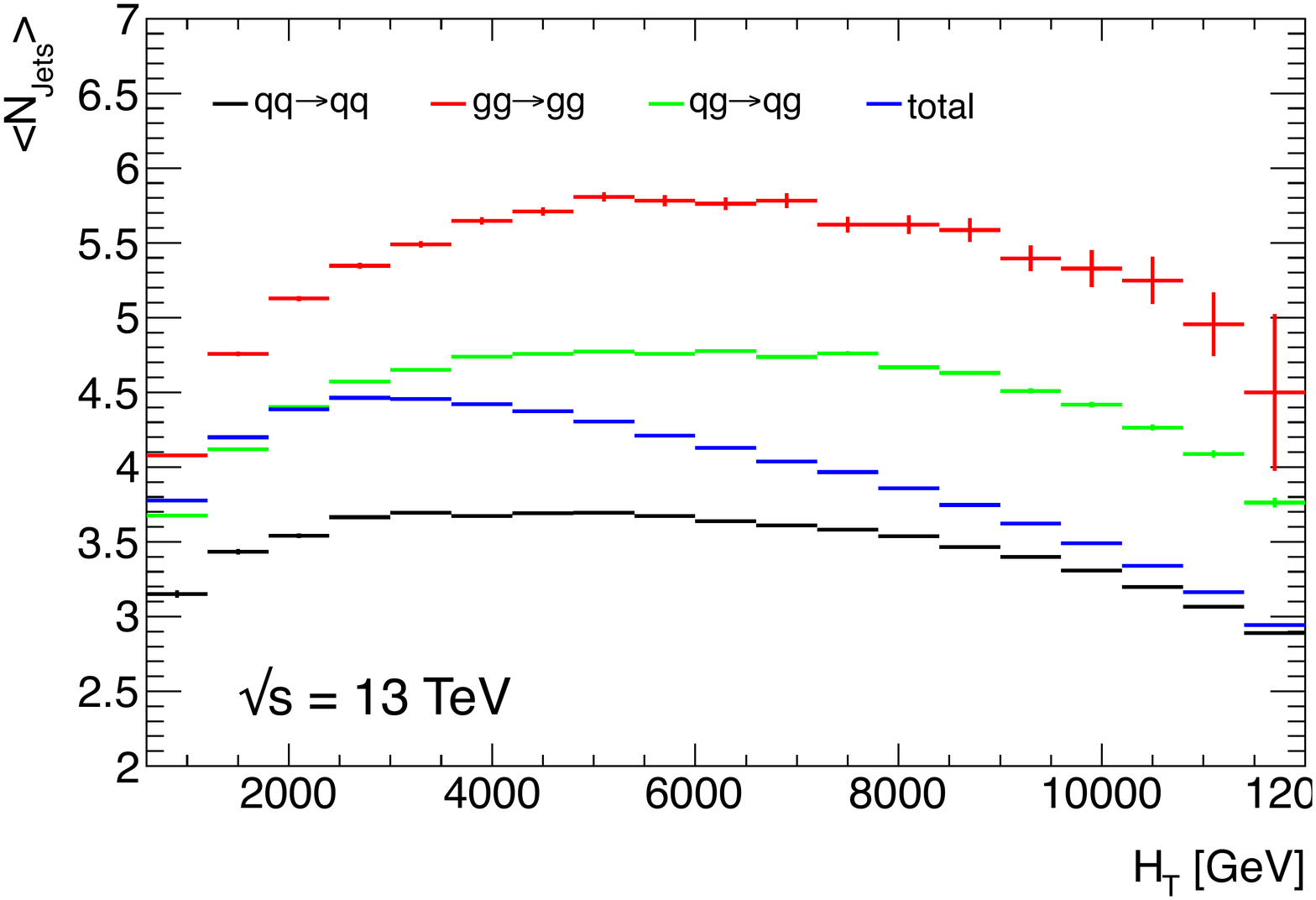}
	\caption{The average number of jets per event (\avnj) as a function of \htt~for the 3 leading subprocesses. Events were generated using Pythia. Note that the presence of gluons in the final state entails higher \avnj. Note also that the \avnj~drop at high values of \htt~appears for all three subprocesses.} 
	\label{NjProc}
\end{figure}

The dependence of the average number of jets per event (\avnj) on the total event's transverse energy as quantified by \htt \footnote{\htt is defined as the scalar sum of the transverse momenta of all the jets in the event, namely: $H_T \equiv \sum_{i=1}^{N_{Jets} }|p_{T_i}|$.}, is shown in Figure \ref{njvsht}. 
One notes the drop of \avnj~ at high values of \htt~exhibited by all the generators that are used in this study.  \avnj~ reaches a maximum at $H_T$ approximately $0.4~E_{beam}$ and then drops by about 25\% at  $H_T$ roughly $1.5~E_{beam}$.
This drop is due, in part, to the drop in the relative cross-sections of subprocesses that contain gluons in their final state (namely, $qg \rightarrow qg$ and $gg \rightarrow gg$) as depicted in Figure \ref{subprocess}. 
Due to their higher  "color"  charge, gluons tend to radiate more jets than quarks (Figure \ref{NjProc}). Therefore,
 a smaller fraction of final state gluons entails lower \avnj.
However, all processes, including $qq \rightarrow qq$, exhibit  the same drop in \avnj~at high \htt, (see Figure \ref{NjProc}) presumably due to the running of $\alpha_s$.

All three generators predict the same dependence of the average jet multiplicity on \htt. However, the absolute value varies by 10\% between the generators in a non-trivial manner:  for example, predictions calculated using the tri-jet NLO method are below the di-jet NLO calculation. The difference is attributed to the fact that the calculations are executed at different perturbative orders and implement different merging schemes and requires further study. In any case, these differences show the need for a data driven approach, as will be presented in the following subsection.


\begin{figure}[!ht]
	\centering
	\begin{subfigure}[c]{0.48\textwidth}
		\includegraphics[trim={1.4cm 0.5cm 0.9cm 0.3cm},clip,width=\linewidth]{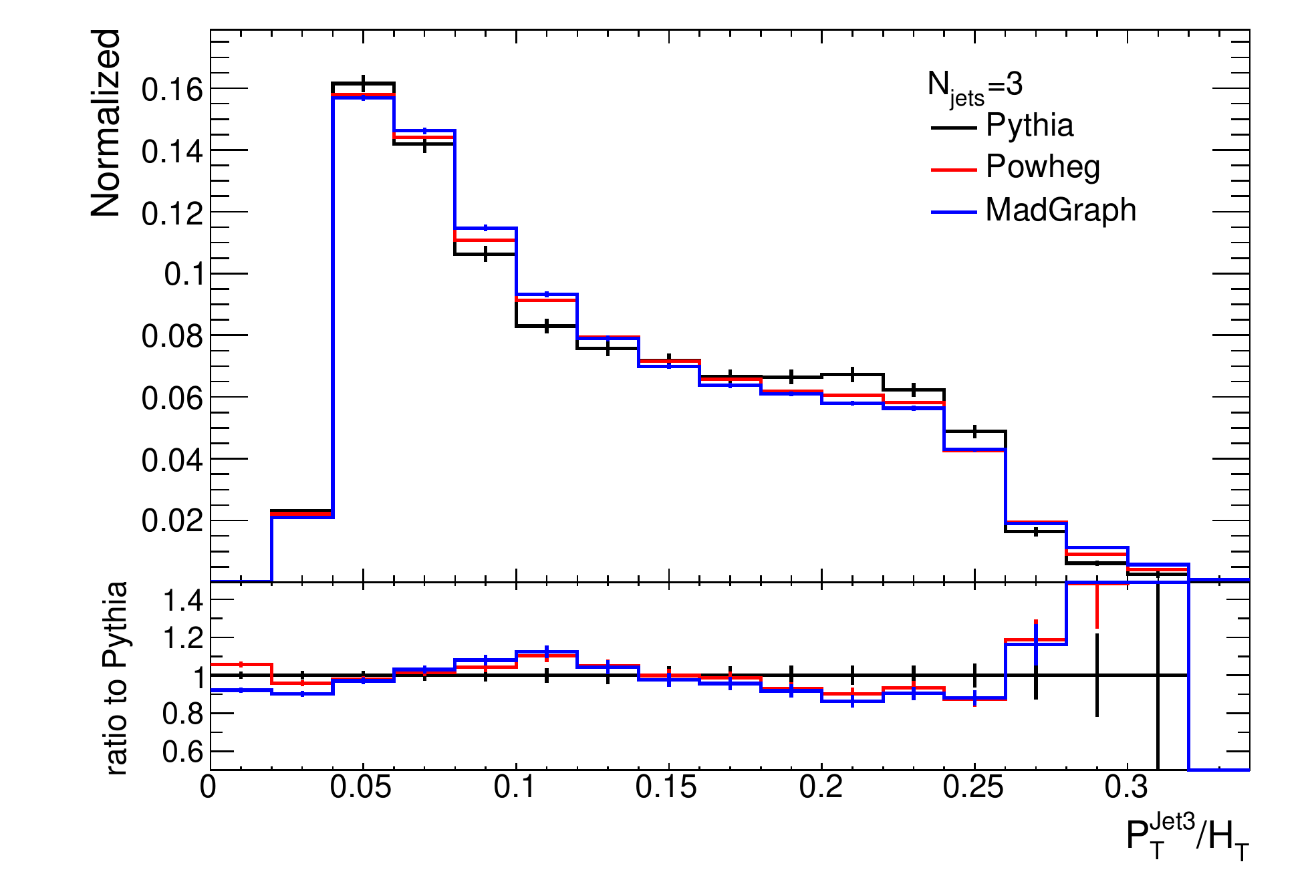}
		\caption{\label{pt3overHT}}
	\end{subfigure}
	\quad
	\begin{subfigure}[c]{0.48\textwidth}
		\includegraphics[trim={1.4cm 0.5cm 0.9cm 0.3cm},clip,width=\linewidth]{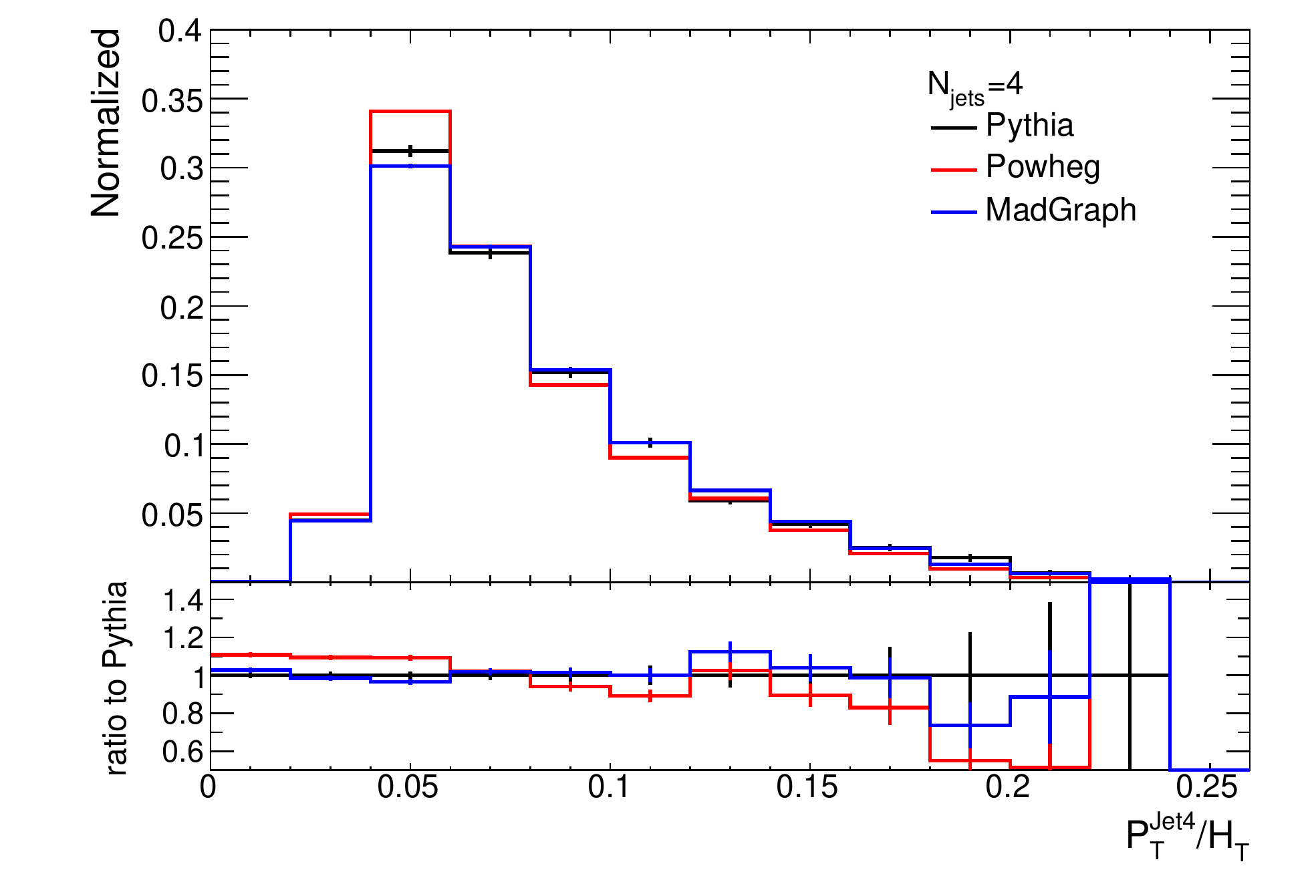}
		\caption{\label{pt4overHT}}
	\end{subfigure}
	
	\caption{ The fraction of the transverse momenta carried out by third jet ($\frac{p_T^{(3)}}{H_T}$) in 3-jets events (left), and by the fourth jet ($\frac{p_T^{(4)}}{H_T}$) in 4-jet events (right). All entries are for events satisfying $H_T > 1$ TeV.}
	\label{fig:ptoverHT}
\end{figure}

As described above,  NLO and multileg calculations are used to generate up to three and four jets respectively. The simulation of higher jet multiplicities is done in all cases using the PS algorithm. 
Therefore, in order to compare the results of the three different simulation strategies the properties of the third and fourth jet (in $p_T$ order) are examined.
 In Figure \ref{fig:ptoverHT}  a comparison of the fraction of the transverse momenta carried out by third jet in events with 3 jets ($\frac{p_T^{(3)}}{H_T}$, top), and by the fourth jet in events with 4 jets ($\frac{p_T^{(4)}}{H_T}$, bottom) are shown.  
One notes that the differences between the three strategies are modest. For 3-jet events Pythia tends to exhibit a small excess of events with high $\frac{p_T^{(3)}}{H_T}$ which is compensated by a lower yield of soft $3^{rd}$ jet. Powheg and Madgraph $\frac{p_T^{(3)}}{H_T}$ distributions look similar.\\

\begin{figure}[!ht]
	\centering
	\begin{subfigure}[c]{0.48\textwidth}
		\includegraphics[trim={1.4cm 0.5cm 0.9cm 0.3cm},clip,width=\linewidth]{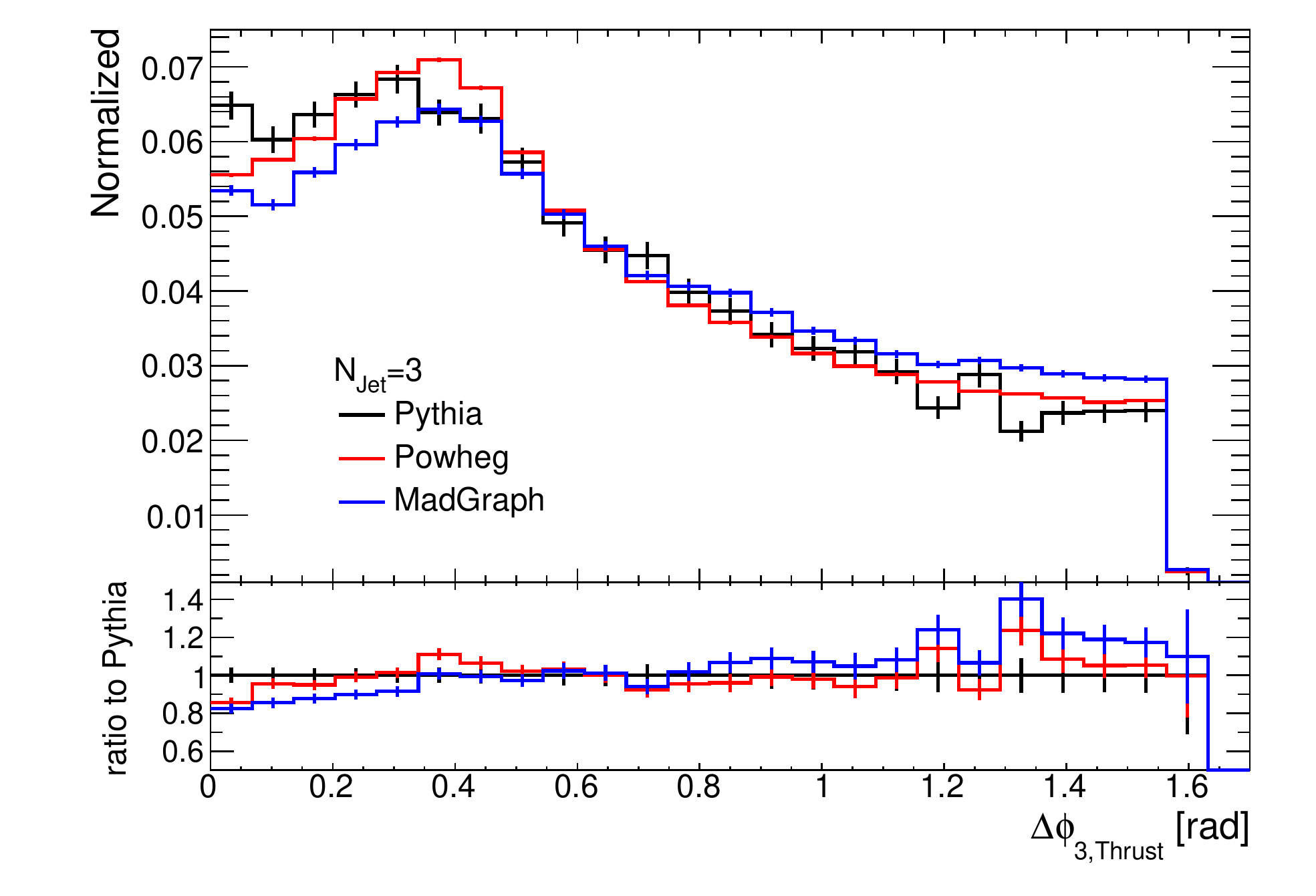}
		\caption{\label{deltaphi3_thrust}}
	\end{subfigure}
	\quad
	\begin{subfigure}[c]{0.48\textwidth}
		\includegraphics[trim={1.4cm 0.5cm 0.9cm 0.3cm},clip,width=\linewidth]{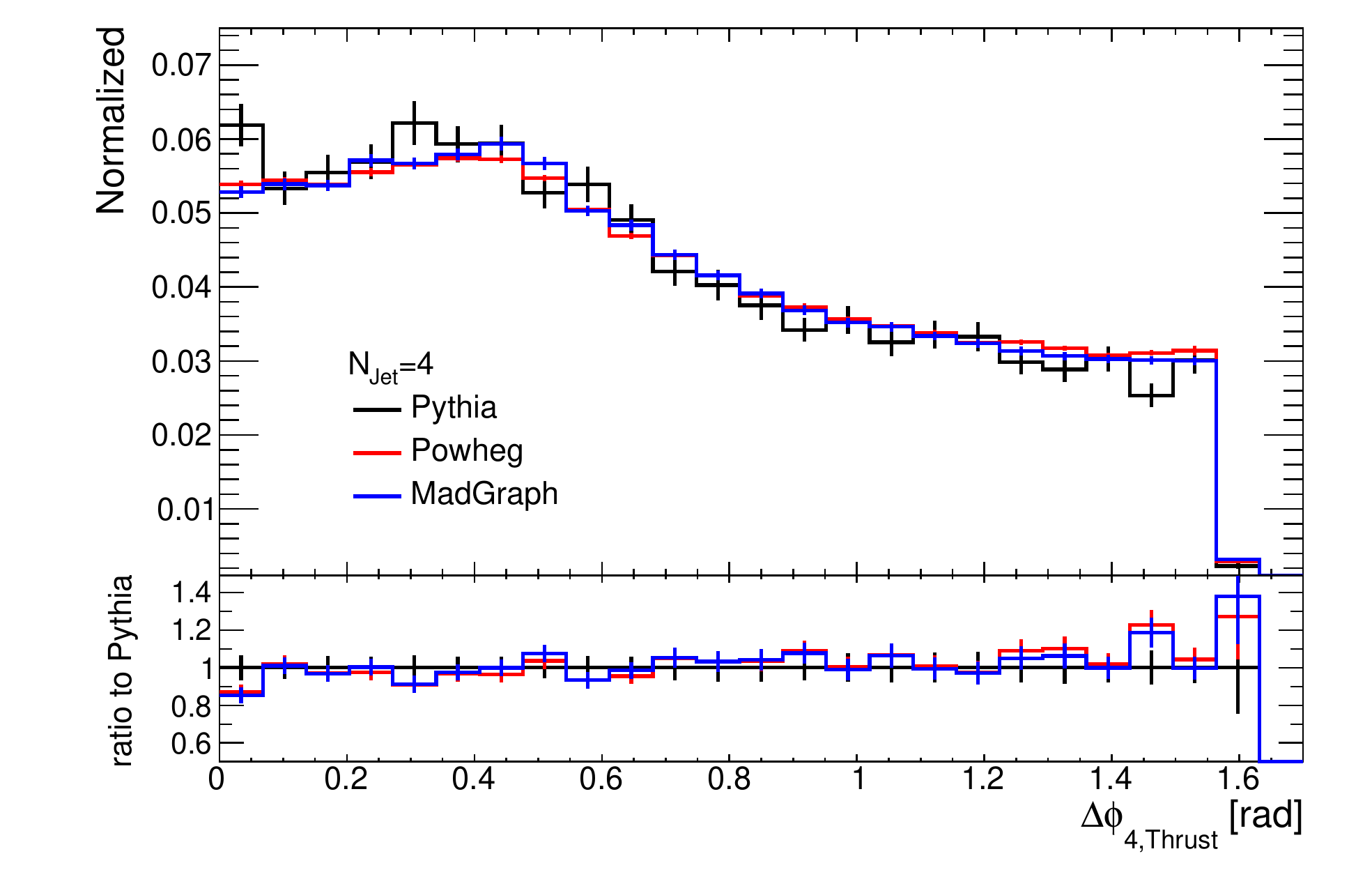}
		\caption{\label{deltaphi4_thrust}}
	\end{subfigure}
	
	\caption{Angular distribution of third jet relative to the transverse thrust axis (see eq. \ref{TthrustPhi}) in 3-jet events (left) and by the fourth jet in 4-jet events (right). All entries are for events satisfying  $H_T > 1$ TeV.
	}
	\label{fig:deltaphi_thrust}
\end{figure}

Similarly, in Figure \ref{fig:deltaphi_thrust} the distribution of the angular separation between the thrust axis and the $3^{rd}$ jet (in 3-jet events, top) and the $4^{th}$ jet (in 4-jets events, bottom) is shown. The transverse thrust axis is defined by:

\begin{equation}
	\mathrm{T_\bot}\equiv \underset{\vec{n}}{\mathrm{max}} \frac{\sum_j \left(\vec{\mathrm{p}}_{\textsc{\tiny{T}}_j}\cdot\vec{n}\right)^2}{\sum_j \vec{\mathrm{p}}_{\textsc{\tiny{T}}_j}^2}
	\label{Tthrust}
\end{equation}
Where $\vec{n}$ is a unit vector and  $\vec{\mathrm{p}}_{\textsc{\tiny{T}}_j}$ is the transverse momentum vector of the $j^{th}$ jet. Using that definition, the azimuthal angle of the Thrust axis ($\phi_\mathrm{T_\bot}$) w.r.t. the $x$ axis of the transverse plane can be evaluated analytically ($j$ index suppressed to avoid cluttering of notation):
\begin{equation}
	\phi_\mathrm{T_\bot} = \frac{1}{2}\arctan\left(\frac{-2\Sigma p_x p_y}{\Sigma\left( p_y^2 - p_x^2\right)}\right) + \kappa\frac{\pi}{2}
	\label{TthrustPhi}
\end{equation}
where:

\begin{equation}
    \kappa =
    \begin{cases}
      1, & \text{if}\ \cos(2\phi_\mathrm{T_\bot})\left(\Sigma p_y^2 - p_x^2\right)<2\sin(2\phi_\mathrm{T_\bot})\Sigma p_x p_y \\
      0, & \text{otherwise}
    \end{cases}
  \end{equation}

In 3-jet events the angular separation between the $3^{rd}$ jet and the thrust axis in Madgraph tends be larger  than that in Pythia, while the same angle in Powheg lies in between. No significant difference between the three simulation strategies is seen for the same distribution in 4-jet events.

\subsection{The Two Hemispheres Method (THm)}
In spite of the reasonable agreement between the outcome of the various QCD event generators, their predictions of  the cross-section and various shape variables of multi-jet events may be at odds with  measurements. Hence, a data-driven procedure for robust modelling of energetic multi-jet events is greatly needed. A new procedure focused on predicting the jet multiplicities is described hereafter. The starting point for this procedure is similar to that taken by ATLAS and CMS in estimating the QCD background for multijet events ~\cite{Khachatryan:2010wx,Mjet_7} , namely, that QCD events can approximately be portrayed as beginning with a \tbt~process that gives rise to two back-to-back (in the x-y plane) out-coming partons, followed by a parton shower. The comparison between LO, NLO and partial NNLO in the previous section shows that the \tbt~picture is not modified significantly by $2 \rightarrow 3$ and $2 \rightarrow 4$ processes.  Multijet QCD events are produced, in this picture, in a sequence of a steps:

\begin{itemize}
\item
The \tbt~matrix element is used and two partons (taken here to be quarks, gluons or a quark and a gluon) are generated with the proper \pt~and $\eta$ distributions;
\item
The \textit{parton shower} algorithm is applied to each parton and secondary partons may be radiated (or split) off the primary ones. The Parton shower is then iteratively applied to the next generation of partons till no more partons are radiated;
\item
Partons are \textit{hadronized} and unstable hadrons are let to decay.
\end{itemize}

In the process's center-of-mass system one can use the plane perpendicular to the initial outgoing partons line of flight to define two hemispheres. In $pp$ collisions the $\hat z$ direction (beam axis) is almost information-free, therefore, projecting the event to the \xy~plane preserves all vital information. The	 line perpendicular to the transverse thrust axis (defined in  eq. \ref{TthrustPhi}) may be used as the dividing line. \\
Because of momentum conservation and the simplistic assumption that the PS is carried out for each of the partons independently, it is claimed here that, at first approximation,  the jet multiplicity in each hemisphere (\njetf~and \njets) are independent of each other. Second order effects (e.g. $qg$ production, ISR etc.) may violate this hypothesis and it will therefor be validated below.

~~     
     
The conjecture, that \textit{\njetf} and  \textit{\njets} \textit{are uncorrelated}, would not be true for a variety of BSM models that give rise to energetic multi-jet events, for example, those mentioned in the introduction. The following procedure is suggested to differentiate between events arising from QCD background and those arising from one of the  hypothetical signals which violates the independence conjecture:
\begin{itemize}
	\item 
	Select \hjm~events having \njetf=1 (i.e. events with one jet in the first hemisphere). The hypothetical \hjm~signal is unlikely to give rise to such events and, therefore, this sample should be  \textit{signal-free} or at least signal depleted.
	\item
	Extract the distribution of the number of jets in the second hemisphere (\njets) from the \textit{signal-free}  (i.e. \njetf=1) sample. This \njets(\njetf=1) distribution should therefore represent the \njets~distribution of pure QCD for all values of \njetf, i.e.  \njetf=2,3,4.., thus serving as a QCD background estimation for those samples which might host signal events.
	\item
	Finally, Compare the \njets~distribution as obtained from the \textit{signal free}  sample with those obtained from the expected \textit{signal region} (\njetf$>1$) distributions. An excess of events with high \njets~ may be considered as a possible indication for the presence of a signal 
\end{itemize}

\begin{figure}[!ht]
\centering
\begin{subfigure}[c]{0.46\textwidth}
  \includegraphics[trim={1.2cm 0.45cm 1cm 0.3cm},clip,width=\linewidth]{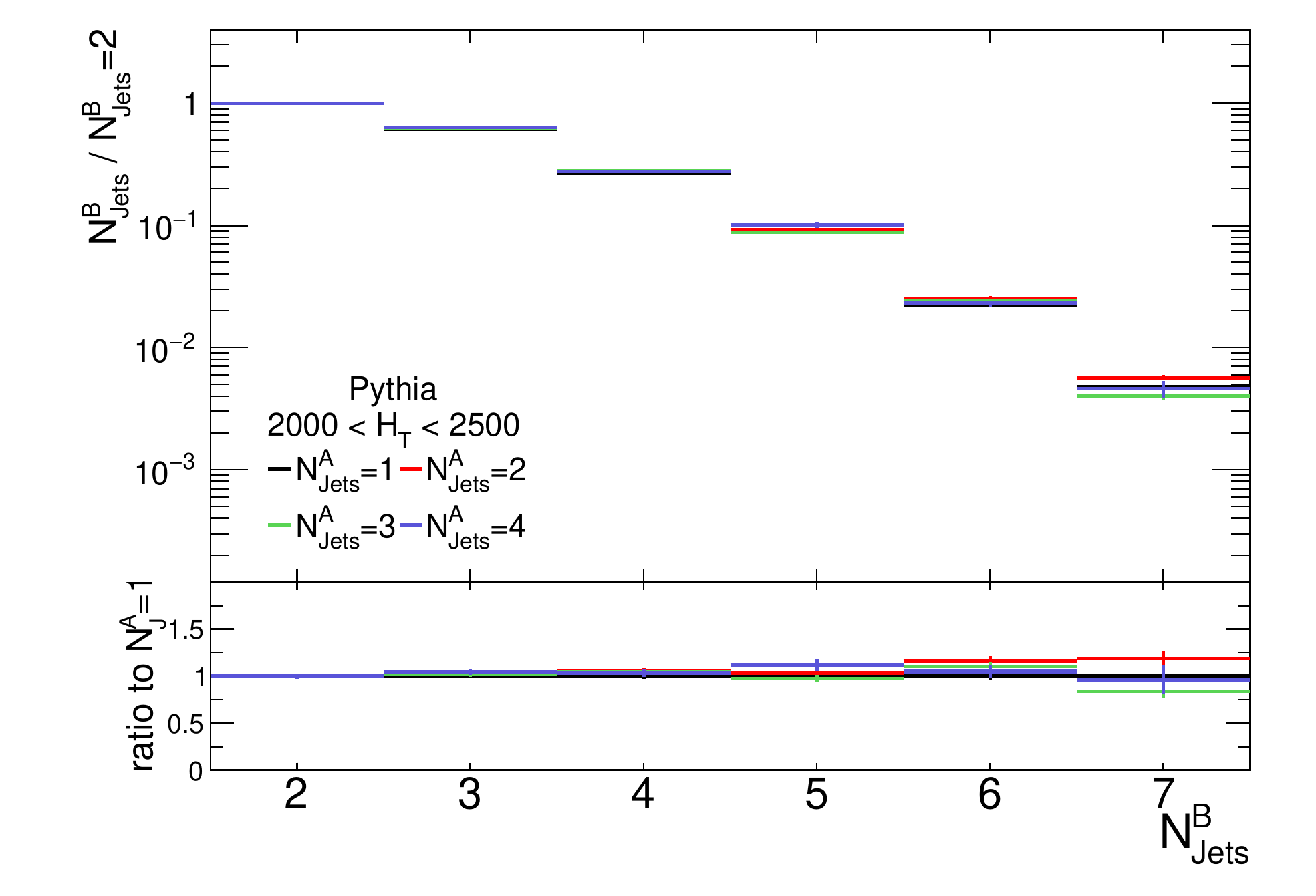}
  \caption{\label{pythia_hemi}}
\end{subfigure}
\quad
\begin{subfigure}[c]{0.485\textwidth}
  \includegraphics[trim={1.2cm 0.45cm 0.9cm 0cm},clip,width=\linewidth]{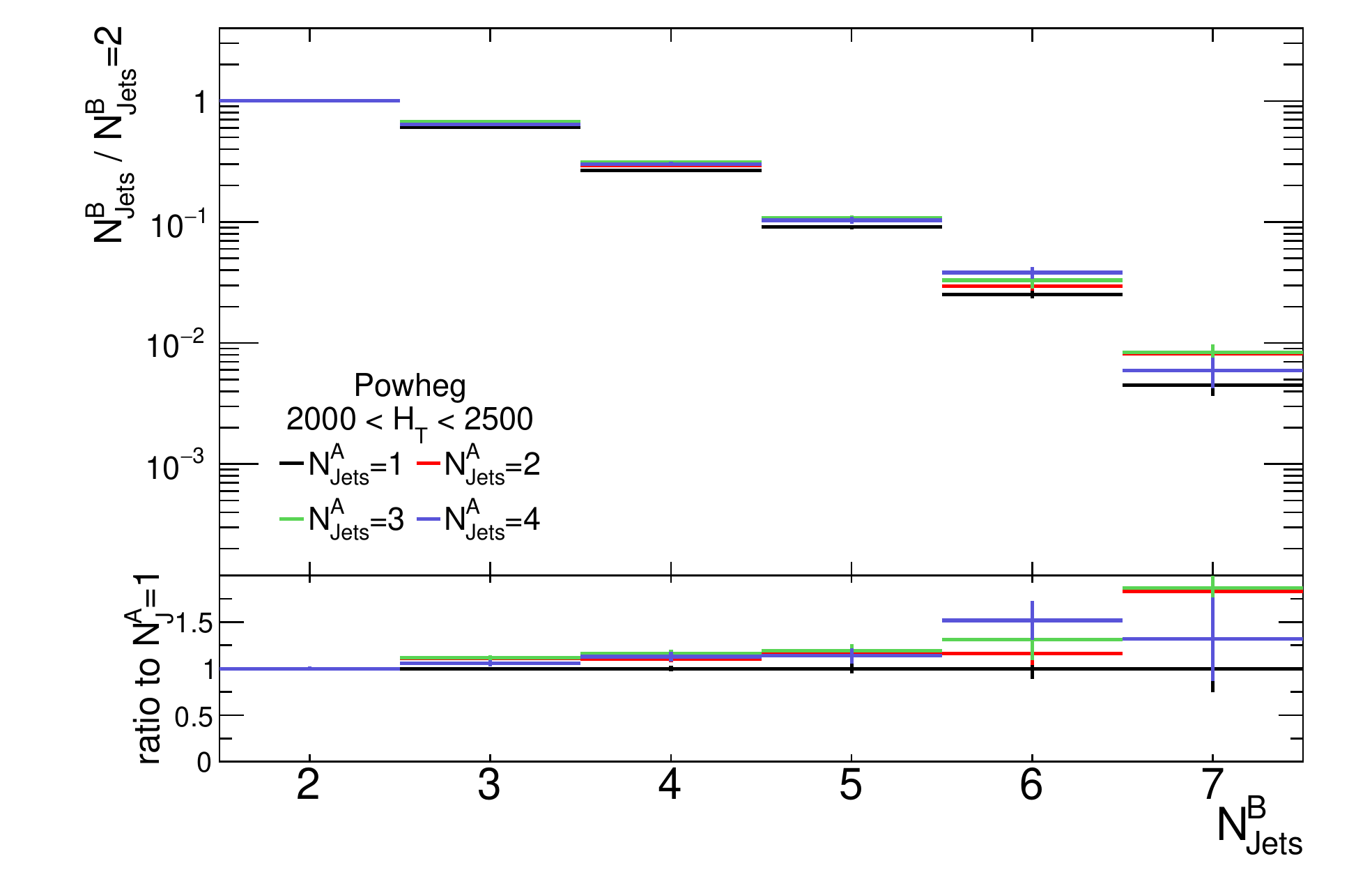}
    \caption{\label{powheg_hemi}}
\end{subfigure}

\quad
\begin{subfigure}[c]{0.48\textwidth}
  \includegraphics[trim={1.2cm 0.45cm 0.9cm 0.3cm},clip,width=\linewidth]{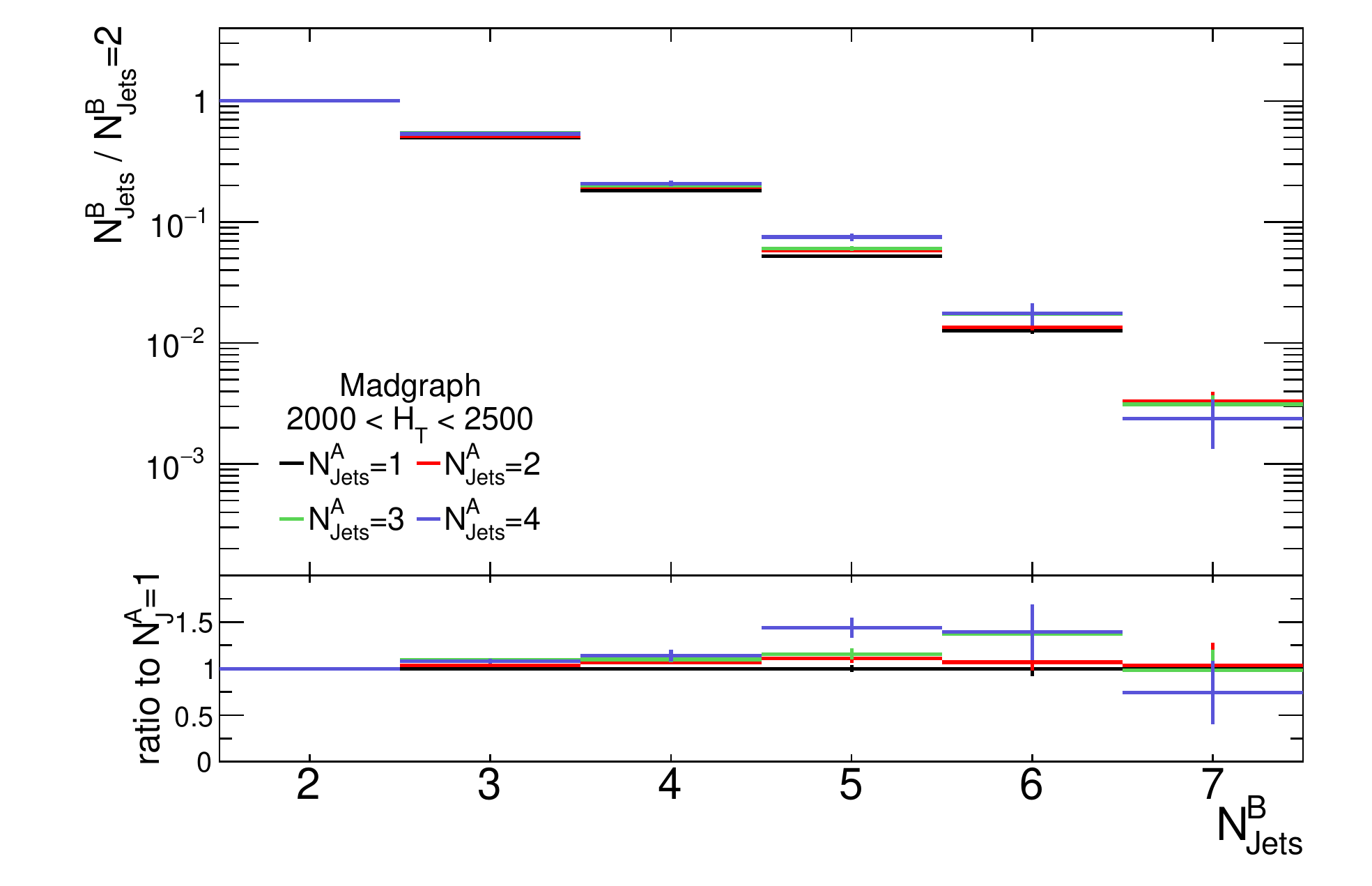}
    \caption{\label{MG_hemi}}
\end{subfigure}

\caption{Hemisphere multiplicity plots showing distribution of \njets~while \njetf~constrained to 1 (black), 2 (red), 3 (green) and 4 (blue). Events were generated by (a) Pythia, (b) Powheg and (c) Madgraph.  Events are selected such that 2$<H_T<$2.5 TeV and all jets has $p_T>$50 GeV and $|\eta|<2.8$. The results are practically independent of these three selection criteria. Events are divided into two hemispheres using the Thrust axis. For visual comparison, all distributions normalized such that content of the \njets=2 bin equals one.}
 \label{fig:hemi}
\end{figure}

As discussed, the above procedure relies on the assumption that for QCD the \njets~ distribution is independent of \njetf. The independence assumption can be tested and validated using QCD simulations by directly comparing \njets(\njetf=1) distribution with those of \njets(\njetf=i) where i=2,3,4.. . Such a comparison is shown in Figure \ref{fig:hemi} (top) using LO events generated by Pythia in the $H_{T}$ region of 2  $< H_T < $ 2.5 TeV. The black markers indicate the distribution of \njets~while \njetf~ is constrained to 1. 
The colored markers indicate the distribution of \njets~while \njetf~is constrained to 2 (red), 3 (green), and 4 (blue). In order to visually facilitate the comparison, all distributions are normalized such that the contents of the second bin (\njets=2) is normalized to one. 
As seen, the \njets~distributions are in good agreement with each other, differing by less than 50\% at the highest bins, which, given the statistical uncertainty, is less than 1$\sigma$.

The simplistic picture of QCD events as basically a $2 \rightarrow 2$ back-to-back events holds only at the LO. The NLO and obviously NNLO or higher orders give rise to more complicated pictures with three, four and more outcoming partons.
Powheg gives rise to $2 \rightarrow 3$ events and Madgraph to $2 \rightarrow 4$. 
Figure \ref{fig:hemi} (center) shows the validity of the independence hypothesis for Powheg and Figure  \ref{fig:hemi} (bottom) shows the same information for MadGraph. As in Pythia, the distributions are in general agreement.

\begin{figure}[h!]
\centering
\includegraphics[width=0.8\textwidth]{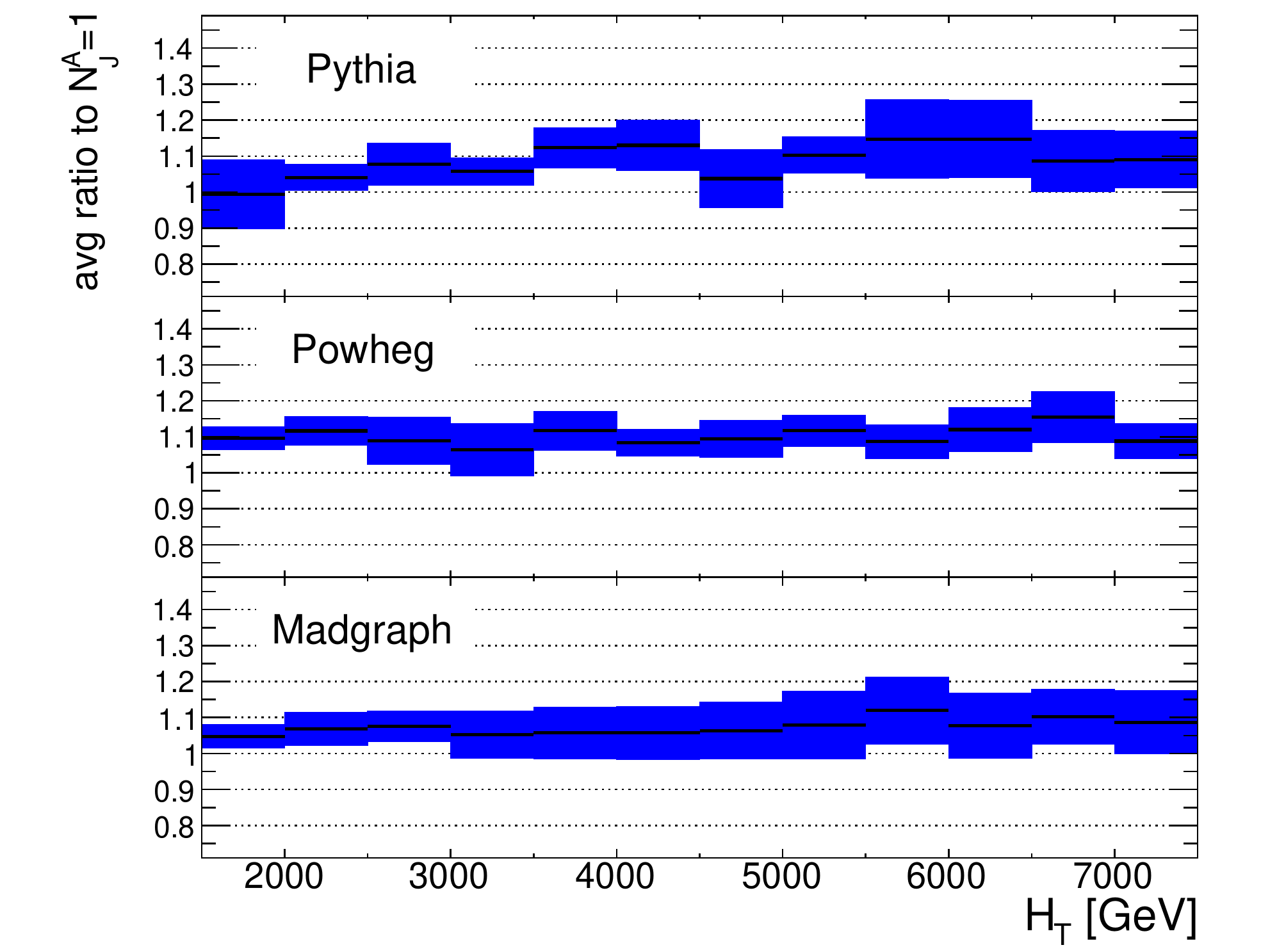}
\caption{Weighted average deviation for each generator for $H_T$ bins of 500 GeV using jet multiplicities of  \njetf=2 through 4 and \njets=3 through 6. Blue error bars mark the weighted standard deviation of the 12 datapoints.} 
\label{offset_weighted}
\end{figure}

A figure of merit for the overall systematic uncertainty of the method, for each $H_T$ range, is evaluated by the average weighted deviation, $\bar{v}$, of the THm prediction compared to the "data" in simulated QCD samples over all \njet~bins in the signal region:

\begin{equation}
\bar{v}(H_T)=\frac{\sum_{j}v_j\frac{1}{\sigma_j^2}}{\sum_{j}\frac{1}{\sigma_j^2}}
\label{weightedavg}
\end{equation}

where the index j runs through the twelve combinations of  \njetf=2 through 4 and \njets~3 through 6,  $v_j$ is the deviation (i.e. ratio of the simulated data to the THm prediction given by \njetf=1) at each datapoint and $\sigma_j$ is the statistical uncertainty. Figure \ref{offset_weighted} summarizes $\bar{v}$ for each generator at different $H_T$ bins of 500 GeV each. Blue error bars mark the weighted standard deviation of the twelve datapoints in each $H_T$ bin, where the weight of a datapoint is defined by its statistical uncertainty.   All deviations, $\bar{v}$, (including errors) are below 25\%.  A precise quantification of the systematic uncertainty to be associated with the THm would be analysis dependant. Additional study may reduce systematic uncertainties.

Other methods currently used in multijet analyses estimate a 1 -- 130\% \cite{Sirunyan:2018xwt} or 40 -- 110\%  \cite{Mjet_13} uncertainty using a fit extrapolation technique  or a 25\% uncertainty using a jet mass template technique \cite{Aaboud:2018lpl}.  Thus the magnitude of the systematic uncertainties of the THm are comparable to those of conventional methods but since the sources of the uncertainties are very different the procedures compliment one another.

\section{Conclusion}
The modeling of jet energy and angular distributions of QCD processes in multi-jet events at 13 TeV was compared for three state-of-the-art MC generation strategies. In particular the kinematics of the $3^{rd}$ jet, mostly affected by NLO and $4^{th}$ jet, mostly affected by NNLO calculations, was studied. The differences between the three models under study were found to be small. 
The average number of jets per event (\avnj) exhibits a drop from approximately 4.5 jets for events with \htt~about 0.4 $E_{beam}$ to roughly 3.7 jets for events with \htt~ close to $1.5~E_{beam}$. Part of this drop is explained by the increase of the relative cross section of $qq  \rightarrow qq$ processes from $\approx$ 25\% of the total cross section at \htt=2.5 TeV to 80\% at 7 TeV at the expense of a drop of the relative cross section of $qg \rightarrow qg$ from 50\% to 20\% and the vanishing of the $gg \rightarrow gg$ process.\\
The comparison  between the different generator predictions of the $3^{rd}$ jet $\textrm{p}_\textrm{T}$ in 3 jet events reveals small differences. Pythia's $3^{rd}$ jet tends to be a bit more energetic than Powheg and Madgraph. the angle that separates this jet from the transverse thrust direction tends to be slightly smaller in Pythia and larger in Madgraph. No significant difference is noticed while studying the properties of the fourth jet. \\
A data-driven procedure for estimating the QCD background for multijet final states, i.e. the \textit{Two Hemisphere Method} (THm) has been proposed.
The basic conjecture of this procedure, namely, the independence of the jet multiplicity in one hemisphere on that in the other, has been tested with the three generators and found to be correct within 25 -- 50\%. A figure of merit estimating the overall systematic uncertainty by including all jet multiplicities for each generator gives a comparable number, approximately 25\% (Figure \ref{offset_weighted}). We consider these results as encouraging. The attainable sensitivity of a THm analysis is comparable to that of the conventional methods. Since the sources of the uncertainties in this new approach are very different from the current methods the procedures compliment one another.

~
~ 
\vspace*{\fill}

\pagebreak
\bibliography{references}

\providecommand{\href}[2]{#2}\begingroup\raggedright\begin{thebibliography}{10}

\bibitem{Giddings:2001bu}
S.~B. Giddings and S.~D. Thomas, \emph{{High-energy colliders as black hole
  factories: The End of short distance physics}},
  \href{http://dx.doi.org/10.1103/PhysRevD.65.056010}{\emph{Phys. Rev.}
  {\bfseries D65} (2002) 056010},
  [\href{https://arxiv.org/abs/hep-ph/0106219}{{\ttfamily hep-ph/0106219}}].

\bibitem{Giddings:2001ih}
S.~B. Giddings, \emph{{Black hole production in TeV scale gravity, and the
  future of high-energy physics}}, {\emph{eConf} {\bfseries C010630} (2001)
  P328}, [\href{https://arxiv.org/abs/hep-ph/0110127}{{\ttfamily
  hep-ph/0110127}}].

\bibitem{Dimopoulos:2001hw}
S.~Dimopoulos and G.~L. Landsberg, \emph{{Black holes at the LHC}},
  \href{http://dx.doi.org/10.1103/PhysRevLett.87.161602}{\emph{Phys. Rev.
  Lett.} {\bfseries 87} (2001) 161602},
  [\href{https://arxiv.org/abs/hep-ph/0106295}{{\ttfamily hep-ph/0106295}}].

\bibitem{Chamblin:2002ad}
A.~Chamblin and G.~C. Nayak, \emph{{Black hole production at CERN LHC: String
  balls and black holes from pp and lead-lead collisions}},
  \href{http://dx.doi.org/10.1103/PhysRevD.66.091901}{\emph{Phys. Rev.}
  {\bfseries D66} (2002) 091901},
  [\href{https://arxiv.org/abs/hep-ph/0206060}{{\ttfamily hep-ph/0206060}}].

\bibitem{Dreiner:1997uz}
H.~K. Dreiner, \emph{{An Introduction to explicit R-parity violation}},
  vol.~21, pp.~565--583.
\newblock 2010.
\newblock \href{https://arxiv.org/abs/hep-ph/9707435}{{\ttfamily
  hep-ph/9707435}}.
\newblock 10.1142/9789814307505\_0017.

\bibitem{Barbier:2004ez}
R.~Barbier et~al., \emph{{R-parity violating supersymmetry}},
  \href{http://dx.doi.org/10.1016/j.physrep.2005.08.006}{\emph{Phys. Rept.}
  {\bfseries 420} (2005) 1--202},
  [\href{https://arxiv.org/abs/hep-ph/0406039}{{\ttfamily hep-ph/0406039}}].

\bibitem{Ellis:2016ast}
J.~Ellis and K.~Sakurai, \emph{{Search for Sphalerons in Proton-Proton
  Collisions}}, \href{http://dx.doi.org/10.1007/JHEP04(2016)086}{\emph{JHEP}
  {\bfseries 04} (2016) 086},
  [\href{https://arxiv.org/abs/1601.03654}{{\ttfamily 1601.03654}}].

\bibitem{Papaefstathiou:2019djz}
A.~Papaefstathiou, S.~Platzer and K.~Sakurai, \emph{{On the phenomenology of
  sphaleron-induced processes at the LHC and beyond}},
  \href{http://dx.doi.org/10.1007/JHEP12(2019)017}{\emph{JHEP} {\bfseries 12}
  (2019) 017}, [\href{https://arxiv.org/abs/1910.04761}{{\ttfamily
  1910.04761}}].

\bibitem{Sirunyan:2018xwt}
{\scshape CMS} collaboration, A.~M. Sirunyan et~al., \emph{{Search for black
  holes and sphalerons in high-multiplicity final states in proton-proton
  collisions at $ \sqrt{s}=13 $ TeV}},
  \href{http://dx.doi.org/10.1007/JHEP11(2018)042}{\emph{JHEP} {\bfseries 11}
  (2018) 042}, [\href{https://arxiv.org/abs/1805.06013}{{\ttfamily
  1805.06013}}].

\bibitem{Mjet_13}
{\scshape ATLAS} collaboration, G.~Aad et~al., \emph{{Search for strong gravity
  in multijet final states produced in pp collisions at $\sqrt{s} =$ 13 TeV
  using the ATLAS detector at the LHC}},
  \href{http://dx.doi.org/10.1007/JHEP03(2016)026}{\emph{JHEP} {\bfseries 03}
  (2016) 026}, [\href{https://arxiv.org/abs/1512.02586}{{\ttfamily
  1512.02586}}].

\bibitem{Aaboud:2018lpl}
{\scshape ATLAS} collaboration, M.~Aaboud et~al., \emph{{Search for
  R-parity-violating supersymmetric particles in multi-jet final states
  produced in $p$-$p$ collisions at $\sqrt{s} =13$ TeV using the ATLAS detector
  at the LHC}},
  \href{http://dx.doi.org/10.1016/j.physletb.2018.08.021}{\emph{Phys. Lett. B}
  {\bfseries 785} (2018) 136--158},
  [\href{https://arxiv.org/abs/1804.03568}{{\ttfamily 1804.03568}}].

\bibitem{Sirunyan:2018adt}
{\scshape CMS} collaboration, A.~M. Sirunyan et~al., \emph{{Event shape
  variables measured using multijet final states in proton-proton collisions at
  $ \sqrt{s}=13 $ TeV}},
  \href{http://dx.doi.org/10.1007/JHEP12(2018)117}{\emph{JHEP} {\bfseries 12}
  (2018) 117}, [\href{https://arxiv.org/abs/1811.00588}{{\ttfamily
  1811.00588}}].

\bibitem{Cacciari:2008gp}
M.~Cacciari, G.~P. Salam and G.~Soyez, \emph{{The anti-$k_t$ jet clustering
  algorithm}},
  \href{http://dx.doi.org/10.1088/1126-6708/2008/04/063}{\emph{JHEP} {\bfseries
  04} (2008) 063}, [\href{https://arxiv.org/abs/0802.1189}{{\ttfamily
  0802.1189}}].

\bibitem{Cacciari:2011ma}
M.~Cacciari, G.~P. Salam and G.~Soyez, \emph{{FastJet User Manual}},
  \href{http://dx.doi.org/10.1140/epjc/s10052-012-1896-2}{\emph{Eur. Phys. J.}
  {\bfseries C72} (2012) 1896},
  [\href{https://arxiv.org/abs/1111.6097}{{\ttfamily 1111.6097}}].

\bibitem{Sjostrand:2014zea}
T.~Sjöstrand, S.~Ask, J.~R. Christiansen, R.~Corke, N.~Desai, P.~Ilten et~al.,
  \emph{{An Introduction to PYTHIA 8.2}},
  \href{http://dx.doi.org/10.1016/j.cpc.2015.01.024}{\emph{Comput. Phys.
  Commun.} {\bfseries 191} (2015) 159--177},
  [\href{https://arxiv.org/abs/1410.3012}{{\ttfamily 1410.3012}}].

\bibitem{Alwall:2014hca}
J.~Alwall, R.~Frederix, S.~Frixione, V.~Hirschi, F.~Maltoni, O.~Mattelaer
  et~al., \emph{{The automated computation of tree-level and next-to-leading
  order differential cross sections, and their matching to parton shower
  simulations}}, \href{http://dx.doi.org/10.1007/JHEP07(2014)079}{\emph{JHEP}
  {\bfseries 07} (2014) 079},
  [\href{https://arxiv.org/abs/1405.0301}{{\ttfamily 1405.0301}}].

\bibitem{Lonnblad:2001iq}
L.~Lonnblad, \emph{{Correcting the color dipole cascade model with fixed order
  matrix elements}},
  \href{http://dx.doi.org/10.1088/1126-6708/2002/05/046}{\emph{JHEP} {\bfseries
  05} (2002) 046}, [\href{https://arxiv.org/abs/hep-ph/0112284}{{\ttfamily
  hep-ph/0112284}}].

\bibitem{Lonnblad:2011xx}
L.~Lonnblad and S.~Prestel, \emph{{Matching Tree-Level Matrix Elements with
  Interleaved Showers}},
  \href{http://dx.doi.org/10.1007/JHEP03(2012)019}{\emph{JHEP} {\bfseries 03}
  (2012) 019}, [\href{https://arxiv.org/abs/1109.4829}{{\ttfamily 1109.4829}}].

\bibitem{Nason:2004rx}
P.~Nason, \emph{{A New method for combining NLO QCD with shower Monte Carlo
  algorithms}},
  \href{http://dx.doi.org/10.1088/1126-6708/2004/11/040}{\emph{JHEP} {\bfseries
  11} (2004) 040}, [\href{https://arxiv.org/abs/hep-ph/0409146}{{\ttfamily
  hep-ph/0409146}}].

\bibitem{Frixione:2007vw}
S.~Frixione, P.~Nason and C.~Oleari, \emph{{Matching NLO QCD computations with
  Parton Shower simulations: the POWHEG method}},
  \href{http://dx.doi.org/10.1088/1126-6708/2007/11/070}{\emph{JHEP} {\bfseries
  11} (2007) 070}, [\href{https://arxiv.org/abs/0709.2092}{{\ttfamily
  0709.2092}}].

\bibitem{Alioli:2010xd}
S.~Alioli, P.~Nason, C.~Oleari and E.~Re, \emph{{A general framework for
  implementing NLO calculations in shower Monte Carlo programs: the POWHEG
  BOX}}, \href{http://dx.doi.org/10.1007/JHEP06(2010)043}{\emph{JHEP}
  {\bfseries 06} (2010) 043},
  [\href{https://arxiv.org/abs/1002.2581}{{\ttfamily 1002.2581}}].

\bibitem{Alioli:2010xa}
S.~Alioli, K.~Hamilton, P.~Nason, C.~Oleari and E.~Re, \emph{{Jet pair
  production in POWHEG}},
  \href{http://dx.doi.org/10.1007/JHEP04(2011)081}{\emph{JHEP} {\bfseries 04}
  (2011) 081}, [\href{https://arxiv.org/abs/1012.3380}{{\ttfamily 1012.3380}}].

\bibitem{Kardos:2014dua}
A.~Kardos, P.~Nason and C.~Oleari, \emph{{Three-jet production in POWHEG}},
  \href{http://dx.doi.org/10.1007/JHEP04(2014)043}{\emph{JHEP} {\bfseries 04}
  (2014) 043}, [\href{https://arxiv.org/abs/1402.4001}{{\ttfamily 1402.4001}}].

\bibitem{Hou:2016sho}
T.-J. Hou et~al., \emph{{Reconstruction of Monte Carlo replicas from Hessian
  parton distributions}},
  \href{http://dx.doi.org/10.1007/JHEP03(2017)099}{\emph{JHEP} {\bfseries 03}
  (2017) 099}, [\href{https://arxiv.org/abs/1607.06066}{{\ttfamily
  1607.06066}}].

\bibitem{Skands:2014pea}
P.~Skands, S.~Carrazza and J.~Rojo, \emph{{Tuning PYTHIA 8.1: the Monash 2013
  Tune}}, \href{http://dx.doi.org/10.1140/epjc/s10052-014-3024-y}{\emph{Eur.
  Phys. J.} {\bfseries C74} (2014) 3024},
  [\href{https://arxiv.org/abs/1404.5630}{{\ttfamily 1404.5630}}].

\bibitem{Khachatryan:2010wx}
{\scshape CMS} collaboration, V.~Khachatryan et~al., \emph{{Search for
  Microscopic Black Hole Signatures at the Large Hadron Collider}},
  \href{http://dx.doi.org/10.1016/j.physletb.2011.02.032}{\emph{Phys. Lett. B}
  {\bfseries 697} (2011) 434--453},
  [\href{https://arxiv.org/abs/1012.3375}{{\ttfamily 1012.3375}}].

\bibitem{Mjet_7}
{\scshape ATLAS} collaboration, G.~Aad et~al., \emph{{Search for low-scale
  gravity signatures in multi-jet final states with the ATLAS detector at
  $\sqrt{s} = 8$ TeV}},
  \href{http://dx.doi.org/10.1007/JHEP07(2015)032}{\emph{JHEP} {\bfseries 07}
  (2015) 032}, [\href{https://arxiv.org/abs/1503.08988}{{\ttfamily
  1503.08988}}].

\end{thebibliography}\endgroup
\bibliographystyle{JHEP}

\end{document}